\documentclass[]{interact}
\pdfoutput=1
\usepackage{subfigure}

\usepackage{amsmath,mathrsfs,amsfonts,bm}

\usepackage{url} 
\usepackage{rotating,threeparttable}

\begin{document}
\title{Variable Selection with Scalable Bootstrap in Generalized Linear Model for Massive Data}
\author{
\name{Zhibing~He\textsuperscript{a}, Yichen~Qin\textsuperscript{b}, Ben-Chang Shia\textsuperscript{c} and Yang~Li\textsuperscript{a,d} \thanks{CONTACT Yang~Li. Email: yang.li@ruc.edu.cn}}
\affil{\textsuperscript{a}Renmin University of China, School of Statistics, Beijing; \textsuperscript{b}University of Cincinnati, Department of Operations, Business Analytics and Information Systems, USA;
\textsuperscript{c}Taipei Medical University, School of Management, Taipei, Taiwan;\textsuperscript{d}Renmin University of China, Center for Applied Statistics}
}
\maketitle

\begin{abstract}
Bootstrap is commonly used as a tool for non-parametric statistical inference to estimate meaningful parameters in Variable Selection Models. However, for massive dataset that has exponential growth rate, the computation of Bootstrap Variable Selection (BootVS) can be a crucial issue. In this paper, we propose the method of Variable Selection with Bag of Little Bootstraps (BLBVS) on General Linear Regression and extend it to Generalized Linear Model for selecting important parameters and assessing the computation efficiency of estimators by analyzing results of multiple bootstrap sub-samples. The proposed method best suits large datasets which have parallel and distributed computing structures. To test the performance of BLBVS, we compare it with BootVS from different aspects via numerical studies. The results of simulations show our method has excellent performance. A real data analysis, Risk Forecast of Credit Cards, is also presented to illustrate the computational superiority of BLBVS on large scale datasets, and the result demonstrates the usefulness and validity of our proposed method.
\end{abstract}


\section{Introduction}

Generalized Linear Model (GLM) and Variable Selection (VS) are two of the popular methods in statistical analysis. The GLM is formulated as a way of unifying various  statistical models, such as logistic regression, and poisson regression, etc. \protect\cite{Miller2010Generalized,nelder1972generalized}.
In GLM, the response, $\mathbf{Y}$, is assumed to be generated from an exponential family distribution, and the mean, $\bm{\mu}$, of the distribution depends on the predictors, $\mathbf{X}$, through:
\begin{equation*}
  E(\mathbf{Y}) = \bm{\mu} = g^{-1}(\mathbf{X} {\bm{\beta}}),
\end{equation*}
where $\bm{ X\beta }$ is a linear combination of unknown parameters $\bm{\beta}$ and predictor $\mathbf{X}$, and $g(\cdot)$ is the link function.
However, the model can be extremely complex when dataset is high-dimensional.
For instance, commercial banks usually build up credit risk prediction model with a large number of attributes  for customers, such as  personal information, and credit records, etc.
In this case, logistic regression is often applied because of its binomial response (i.e. 0 for without risk and 1 for with risk),
but it may involve insignificant predictors and result in introducing redundant errors into the model. To find out the attributes which actually affect the response on the interpretation of predictive study, variable selection modeling based on penalization is used as one of the common methods \protect\cite{cantoni2005variable,pourahmadi2011covariance}.

In order to assess the quality of estimators for variable selection, non-parametric bootstrap, which has good theoretical properties to evaluate basis and quantify uncertainty of estimates (e.g. via a standard error or a confidence interval), is popularly used.
However, in the era of Big Data, datasets of  massive size has become increasingly prevalent, and bootstrap is often realized as a substantial expense.
Since the samples drawn from bootstrap have the same order size of original data, the use of bootstrap method becomes severely blunted in the large datasets.
Thus, even one simple estimation could be computationally demanding, and the repeated estimations on the comparably large sized resamples would be prohibitively intensive. To mitigate this problem, a natural solution is to exploit the modern trend toward parallel and distributed computing.
Indeed,  bootstrap would seem ideally suited to this by using different processors or compute nodes to process different bootstrap resamples independently in parallel.
However, the large size of bootstrap resamples in the massive data setting renders this approach problematic, This is because the cost of indepenent computing resource on a single resample can be overly high.
With respect to the study of risk prediction for a commercial bank, the data delivered to each independent processor could reach 7GB when the original dataset has approximately 5 million data points and 25 categorical predictors.
Rather obviously, the computation for traditional Variable Selection with Bootstrap (BootVS) would be extremely costly.

Motivated by the need of an accurate but scalable method for estimating parameters and  assessing performance in large datasets, especially under  the  situation of GLM, we propose the Variable Selection with Bag of Little Bootstraps (BLBVS). The method is inspired by the idea of Bag of Little Bootstraps (BLB) \protect\cite{kleiner2014scalable} which bootstraps multiple smaller subsets of a larger dataset, and then incorporates it with the method of variable selection.
That is, instead of resampling the original large dataset, the method of BLB is employed to bootstrap to bootstrap subsets of reduced size and then apply variable selection as the resampling level.
Therefore, the computational cost is reduced since it is proportional to the subset size. BLBVS is also more efficient on computational profile than BootVS, as it only requires estimations under the consideration of many but much smaller sub-datasets.
We have shown that BLBVS can convergence to a relative low standard error with higher accuracy faster and faster speed than the BootVS through numerical studies.

The remainder of this paper is organized as follows.  In  Section 2, we introduce the method of BLBVS for Generalized Linear Regression (GLR) by using Lasso and then extend to GLM by using Group Lasso. Section 3 shows that  BLBVS has faster convergence with higher accuracy compared with traditional BootVS. The scalability and computation analysis is given in Section 4.  A real data analysis of  risk forecast for a commercial bank is presented in Section 5 to explore the performance of our method. Section 6 discusses out conclusion and potential future areas of focus.

\section{Methodology}
\subsection{Method: Variable Selection with Bag of Little Bootstraps (BLBVS) for GLR}

Variable selection is well known for its property of selecting a subset of relevant predictors to the response for the learning model \protect\cite{Fan2001Variable}.
General Linear Regression consists of  a continuous response $\mathbf{Y}\in R^n$, a $n\times p$ design matrix $\mathbf{X}$ and a parameter vector $\bm{\beta} \in R^p$,
which is referred to get a general form of variable selection. A subset of $\beta$ can be solved by minimizing the following function:
\begin{equation}\label{eq:glr}
  S_{\lambda}(\bm{\beta})=||\bold{Y}-\bold{X}\bm{\beta}||_2^2 +\lambda \times P(\bm{\beta}),
\end{equation}

where $||\bm{u}||_2^2=\sum_{i=1}^{n} u_i^2$ is defined for a vector $\bm{u}\in \mathbb{R}^n$, and $\lambda$ is the tuning parameter which controls the severity of constraints on the regression model. $\lambda$ is chosen by cross-validation in order to minimize overall error rate.
In Equation \eqref{eq:glr}, with increasing of $\lambda$, the penalization is getting intensified and fewer variables would be selected.
Our primary variable selection model is the Regression with Lasso penalty \protect\cite{tibshirani1996regression}:
\begin{equation}\label{eq:lasso}
\begin{split}
& \mathop {\text{minimize}}_{\beta}{||\mathbf{Y}-\mathbf{X}\bm{\beta}||^2_2} \\
& \text{subject to} ||\bm{\beta}||_{1}\leq t. \\
\end{split}
\end{equation}
It can be rewriten in the Lagrangian form:
\[\mathop{\text{minimize}}_{\bm{\beta} \in \mathbb{R}^p}{||\mathbf{Y}-\mathbf{X}\bm{\beta}||^2_2 +\lambda||\bm{\beta}||_1}.  \]
A subset of revelant parameters can be derived by applying Cyclic Coordinate Descent (CCD) algorithm in equation \eqref{eq:lasso}.
It is important to note that many other penalty functions are available and can be used with variable selection methods, such as SCAD  \protect\cite{Fan2001Variable,yang2011efficient}, Elastic Net  \protect\cite{Zou2005Regularization,de2009elastic,zou2005addendum}, Adaptive Lasso  \protect\cite{Zou2006The,zhang2007adaptive}, Group Lasso  \protect\cite{yuan2006model,meier2008group}, MCP  \protect\cite{Zhang2010Nearly,jiang2013cross} etc.

In addition to selecting and  estimating the parameters, it is always of interest to explore the uncertainty of the estimation.
To assess the quality of estimators in variable selection model, and to reduce the computational cost of BootVS on massive dataset (as mentioned in Section 1), BLBVS is introduced.
The workflow of BLBVS is shown in Figure \ref{BLBVS workflow}.
Suppose the observed data ($x_1,\cdots,x_n$) is with $n$ data points which are identically and  independently distributed,
randomly select $s$ subsets (which are also called bags or modules) of smaller size $\mathit{b} = \{\mathit{n}^\gamma| \gamma \in (0, 1)\}$  without replacement.
For each subset,  bootstrap $r$ resamples with replacement to make each of them  has size $n$ (same size as original data), and thus each resample contains at most $b$ distinct data points.
The distribution of each resample  is realized by assigning a random weight vector $n^{*}=(n_1^*,\cdots,n_b^{*})$ over the $b$ distinct data points of the corresponding subset: $\textup{Multinomial}(n, (1/b){1}_b)$, where $1_{b}$ is the unit vector of length  $b$ and  the weights are subject to $\sum\limits_{i = 1}^b {n_i^*}= n$.

Let $\hat{\beta}_{ij}$ be a vector of estimators in the $j$th resample of $i$th subset and $\hat{\xi}_i$ be the standard error of parameter estimation in $i$th subset.
In our proposed method, the overall standard error $\hat{\xi}$ of parameter estimation  of the original dataset is calculated by averaging the results of multiple subsets $\hat{\xi}_i\ i=1,\cdots,s$, according to BLB.
With regard to variable selection, an indicator function $\mathbf{I}(\hat{\beta} \neq 0)$ is generated to represent the selection result for each vector of predictors $\hat{\beta}_{ij}$ for per resample per subset.
To drive the final result of variable selection for the original dataset, a voted criterion similar to the theories of Decision Trees  \protect\cite{safavian1990survey,kohavi1996scaling} and Random Forest  \protect\cite{breiman2001random,liaw2002classification} is applied.
Suppose each resample in the corresponding subset has the same importance, denoted as one vote, then the total number of votes is $s \times r$.
The predictors which get the majority votes would be selected, and thus the finial selection proportion $p$ of each predictor is defined as
\begin{equation}\label{eq:prop}
p=\frac{\sum_{i=1}^{s}\sum_{j=1}^{r}\mathbf{I}(\hat{\beta}(X_{ij}^{*})\neq 0)}{s\times r}.
\end{equation}
The predictor would be selected if it satisfies $p>c$, where the cut-off $c$ can be determined by the definition of majority in different studies.
\begin{figure}[!htbp]
\centering
\includegraphics[width=0.7\textwidth]{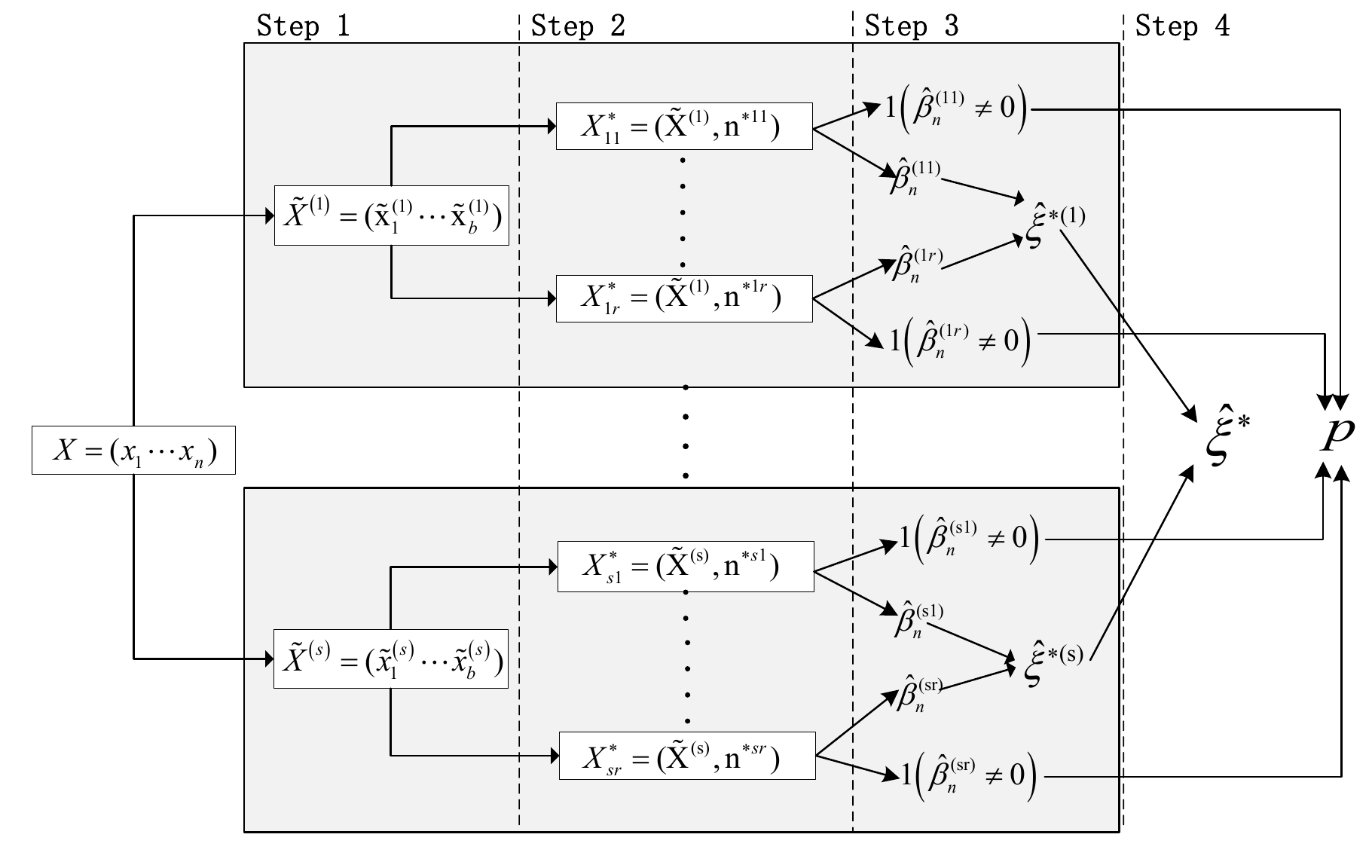}
\caption{BLBVS workflow:  the original massive dataset has $n$ data points; (Step 1) exclusive subsets of size $b\ (b<n)$ are sampled randomly without replacement;
(Step 2) $r$ resamples of size $n$ are drawn from $b$ dsitinct objects with replacement; (Step 3) estimators and their corresponding indicators are calculated per resample per subset, the standard error of parameter estimation is computed by incorporating the estimation results of resamples at the subset level; (Step 4) the overall standard error is derived by taking the average across subsets, and a suggested set of parameters are obtained by getting the selection proportion $p$ and choosing those with the most votes.
}\label{fig:BLBVS workflow}
\end{figure}

Unlike computing $n$ data points from the original dataset, the resamples bootstrapped from each subset only contain at most $b$ distinct objects, therefore results in much less computational consumption.
As mentioned by  \protect\cite{Efron1993An}, when the dataset size is very large, the number of distinct points in each resample could be simply computed by formula $0.632n$
That is, when the original dataset has size $\mathit{n}= 1,000,000$,each subset   approximately has size $\mathit{b} = 3,981$ if $\gamma = 0.6$.
The difference between BLBVS and BootVS could be ideally described by storage occupation: if we assume each data point occupies 1MB of storage space, then the original dataset would occupy 1TB, a conventional bootstrap resample would occupy approximately 632GB, but a BLBVS subset or resample only needs at most 4GB of storage.
It is pretty obvious that our proposed method requires much less computing resource even when the original dataset is extremely large.

The pseudo-code of BLBVS is shown in Table \ref{table:algorithm}.
According to \protect\cite{kleiner2014scalable}, it is suggested to set $\gamma \in [0.6,0.9]$.

\begin{table*}[!htbp]
\small
\centering
\caption {Algorithm of Variable Selection with Bag of Little Bootstraps (BLBVS)}\label{table:algorithm}
\begin{tabular}{l }

\hline
\textbf{Algorithm 1}: Bag of Little Bootstraps Variable Selection (BLBVS)\\
\hline
\textbf{Input:} Original Data $X=(x_1,\dots,x_n)$\\
$b$: size of each subset\\
$s$: number of subsets\\
$r$: number of resamples/ number of Monte Carlo iterations\\
${\beta}$: parameter vector\\
$\xi$: assessment of estimator quality/ standard error of estimator \\
\textbf{Output:}\quad  an estimate of $\xi$ and the selection result of proportion $p$ \\
\; 1. \textbf{for}$ i = 1$ to $s$ \textbf{do}\\
\; 2.\quad Randomly select a subset $\mathcal{I}=\{i_1,\cdots,i_b\}$ of $b$ objects from $\{1,2,\cdots,n\}$\\
\ \qquad  without replacement.\\
\; 3.\quad Form the subset $\tilde{X}^{(i)}$ based on the original dataset $X$ and index set $\mathcal{I}$ \\
\; 4.\quad \textbf{for}$ j = 1$ to $r$ \textbf{do} \\
\; 5.\quad\quad Generate $r$ bootstrap resamples: randomly draw a resample $X^{*}_{ij}=(\tilde{X}^{(i)}; n^{*ij})$  \\
\ \quad\qquad   of size $n$ from subset ${\tilde{X}}_i$  with replacement,  where each resample is \\
\ \quad\qquad  \emph{Multinomial}$(n, (1/b)\mathbf{1}_b)$ distributed with a weight vector on $b$ distinct data points.\\
\; 6.\quad\quad VS model is applied on resample $X^{*}_{ij}$ to estimate the vector of parameters ${\beta}$ and \\
\ \quad\qquad calculate $\mathbf{1}(\hat{\beta}(X_{ij}^{*})$. \\
\; 7.\quad \textbf{end} \textbf{for}\\
\; 8.\quad Get $r$ estimated results of parameter vector $\beta$ and then calculate the standard\\
\ \quad\quad  error of parameter estimation $\xi^{*(i)}$.\\
\; 9. {\textbf{end} \textbf{for}}\\
10. Compute the finial estimate of $\xi$ by $s^{-1}\sum_{i=1}^{s}\xi_i^{*(i)}$\\
11. Return the proportion $p$ based on Eq.\eqref{eq:prop}\\
\hline
\end{tabular}
\end{table*}

\subsection{Extension: BLBVS
 with Group Lasso Penalty for GLM}

The variable selection method discussed in last section is not suitable for nominal or ordinal predictors selection. This is because it may cause a result of choosing partial dummy predictors within the same categorical variable, i.e. choosing some dummy predictors but  abandon the rest.
More, the selection results may vary for different reference level settings \protect\cite{li2013grouplasso}. It leads to confusion because the inconsistent results for the same dataset. For the convenience of interpretation, all the dummy variables that transformed from the same categorical predictors should be selected or not be selected at the same time. In other words, if a categorical predictor contains $k$ categories, $k-1$ transformed dummy predictors should be selected or abandoned together as a group. Therefore a suitable extension of Lasso, Group Lasso \protect\cite{meier2008group} is proposed to overcome the aforementioned issues by grouping all dummy codes per categorical predictor and performing variable selection at the group level. An attractive property of this is the invariant characteristic under orthogonal transformations like ridge regression \protect\cite{yuan2006model}. The estimator of Group Lasso is defined as
\begin{equation}
\hat{\bm{\beta}}_{\lambda}= \text{arg} \min(||\mathbf{Y}-\mathbf{X}{\bm{\beta}}||^2_2+{\lambda}\sum_{g=1}^{G}||\bm{\beta}\mathcal{I}_g||_2)
\end{equation}
where $\mathcal{I}_g$ is the index set belonging to the $g$th group of variables, $g=1,\dots,G$.

Suppose there is a binary response $Y_i$ and $p$ independently and identically distributed predictors $X_i$, including both continuous and categorical predictors.
The predictors can be divided into G groups by considering all dummy variables transformed from the same categorical predictor as one group with degree of freedom $df_g=k-1$ (where $k$ is the number of categories), and considering each continuous predictor has $df_g=1$.
$X_i$ can be then  rewritten as $(X_{i,1},\cdots,X_{i,G})$, where $X_{i,g}$ stands for predictors which belong to group $g$.
For example, three dummy variables that transformed from a categorical predictor of four level will be treated as one group with $df=3$.
Let $\bm{\beta}_g$ be the parameter vector corresponding to the $g$th group.
\begin{equation}
  log\left(\frac{p_\beta(\mathbf{X})}{1-p_\beta(\mathbf{X})}\right)=\eta_{\beta}(\mathbf{X})=\beta_0+\sum_{g=1}^{G}X_{i,g}^{T}\bm{\beta}_g,
\end{equation}
where $p_{\beta}(x)=P_{\beta}(Y=1|x)$ is the conditional probability.  $\beta_0$ is the intercept, and $\bm{\beta}_g \in \mathbb{R}^{df_g}$ is the parameter vector corresponding to the $g$th group.   The whole parameter vector is denoted as $\beta \in \mathbb{R}^{p+1}$, i.e. $\bm{\beta} =(\beta_0,\beta_1^{T},\cdots,\beta_G^{T})^T$, it could be estimated by minimizing the following convex function \protect\cite{meier2008group}:
\begin{equation*}
  S_{\lambda}(\bm{\beta})=-l(\bm{\beta})+\lambda \sum_{g=1}^{G}s(df_g)\|\bm{\beta}_g\|_2,
\end{equation*}
where $l(\cdot)$ is the log-likelihood function, $l(\beta)=\sum_{i=1}^{n}y_i\eta_{\beta}(x)-log[1+exp\{\eta_{\beta}(x)\}]$.
$s(\cdot)$ is used to rescale the penalty and is often related to the dimension of parameter $\bm{\beta}_g$.
Usually, the value of $\lambda$ is chosen by cross-validation to minimize the overall error rate.

\section{Simulations}
\label{sec:simu}

In this section, statistical performance of BLBVS is explored by comparing correctness of selection result and the convergence properties  with the traditional BootVS method.
All experiments in this section are implemented and executed using software R \url{(http://www.r-project.org/)} on a single processor (Windows 7 system; Inter i5-3230M and 2.6GHz CPU; 4GB RAM).

Logistic regression with continuous independent predictors is considered as our true model and Group Lasso is used as our penalty function..
The simulated data drawn from the underlying distribution $\emph{P}$ are identically and independently distributed and has the form $(\tilde{\mathbf{X}}_i,Y_i)$ for $i =1,\cdots,n$, where $\tilde{\mathbf{X}}_i \in \mathbb{R}^{d}$ and $Y_i \in \{0,1\}$.
Let $\mathcal{I}_T$ denote index of grouped variables in the true model, i.e.  $\mathcal{I}_g \in \mathcal{I}_T$ stands for the $g$th group of variables.
$\hat{\bm{\beta}}_g$ is the parameter estimate vector in  $\mathbb{R}^{d}$ and $\hat{\xi}$ is the standard error of $\hat{\bm{\beta}}_g$.
In this paper, we perform simulation by considering the formula $b=n^{\gamma}$ with  $\gamma \in \{0.6,0.7,0.8\}$ and let $r=100$ in each sub set.
For both methods, identical evaluation criteria are used to evaluate the results of selected variables and quality of estimation.

To simulate dataset of a true underlying distribution $\emph{P}$ in logistic model: we set the group $G=8$, data size $n=20,000$  and 35 continual independent variables (variable allocation is shown in Table \ref{Tab:allo}). $\tilde{X}_i$ is drawn independently from the normal  distribution: $\tilde{X}_{i,g} \sim \text{Normal}(0,1)$, and $Y_i$ is drawn from the Bernoulli distribution: $Bernoulli(1,p_{\beta}(x))$, where
\begin{equation}
  p_{\beta}(x)= \{1+exp(-\tilde{X}_{i,G}^{T}\bm{\beta_g})\}^{-1}.
\end{equation}
Assume five groups of variables are selected and thus indicated in the index vector  $\mathcal{I}_T = \{\mathcal{I}_1, \mathcal{I}_2, \mathcal{I}_4, \mathcal{I}_6, \mathcal{I}_7 \}$.
If $\mathcal{I}_g \in \mathcal{I}_T$ , set the true value of corresponding parameter $\bm{\beta}_g$ is 10.
Otherwise, $\bm{\beta}_g$ is randomly drawn from the distribution $\text{Normal}(0,1)$, which is intended to be shrunk to $0$ in variable selection models.
\begin{table}
\centering
\caption{Grouping situation of 35 continuous predictors}\label{Tab:allo}
\begin{tabular}{cc}
\hline
index set & index variables\\
\hline
$\mathcal{I}_1$ & \{1,$\cdots$,5\} \\
$\mathcal{I}_2$ & \{6,$\cdots$,9\} \\
$\mathcal{I}_3$ & \{10,$\cdots$,15\} \\
$\mathcal{I}_4$ & \{16,$\cdots$,20\} \\
$\mathcal{I}_5$ & \{21,$\cdots$,25\} \\
$\mathcal{I}_6$ & \{26,$\cdots$,28\} \\
$\mathcal{I}_7$ & \{29,$\cdots$,31\} \\
$\mathcal{I}_8$ & \{32,$\cdots$,35\} \\
\hline
\end{tabular}
\end{table}

To evaluate the assessment procedures on a given estimation and true underlying data distribution $P$,
firstly we should compute the ground truth $\xi(Q_n(P))$ by generating 1,000 realizations of original  datasets from the true underlying distribution, then computing $\hat{\theta}_n$ on each realization and using this collection of $\hat{\theta}_n$'s to form an approximation to $Q_n(P)$.
For an independent dataset realization of size $n$ from the true underlying distribution, we run each quality assessment procedure until it converges and record the estimate of $\xi(Q_n(P))$.

\subsection{Correctness of selection}

In the above setting, we set $\mathcal{I}_T$ includes $\mathcal{I}_{1,2,4,6,7}$ , namely these groups of variables are in the true model and they should be selected.  The criterion of selection is defined in Eq.\eqref{eq:prop}.

\begin{figure}[!t]
\begin{center}
\includegraphics[width=0.6\textwidth]{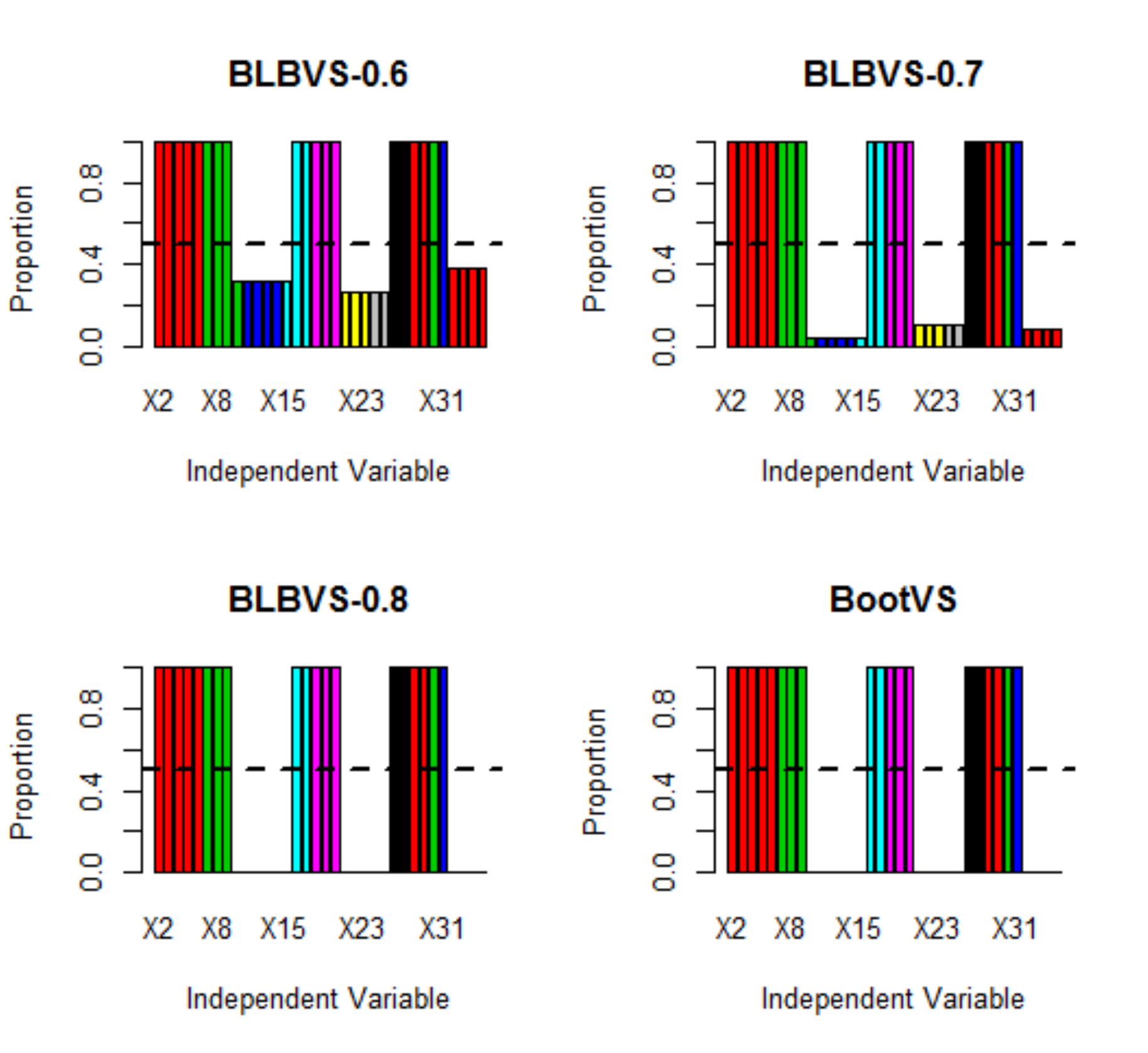}
\end{center}
\caption{Results of variable selection for different $\gamma$.\ Note,The independent variables from the same group have the same color. In the true model, The selection results in all full situations matching our predefined true modell, where $\mathcal{I}_T=\mathcal{I}_{1,2,4,6,7}$.}\label{fig:sele}
\end{figure}

Figure \ref{fig:sele} shows the selection results for BLBVS　with various values of $\gamma$ and BootVS. It is obvious that the final selections are the same and completely correct in both methods.
For BLBVS, our results show the proportion becomes more accurate as $\gamma$ increases. We can also see the performance of BootVS is superior to that of BLBVS, especial when $\gamma$ is relatively small. This is because the proportion is computed base on the information of resample, so it would be more accurate if the resample contains more distinct data points.
From previous knowledge, we know there are almost $0.632n$ different data points in each resample of BootVS, while only $n^{\gamma}$ in each running of BLBVS.  It makes sense that the selection superiority of BootVS reduces as $\gamma$ grows, we could see  that the results are almost the same for both methods when $\gamma = 0.8$.
Meanwhile, we construct the empirical confidence interval for each coefficients.
It is not difficult to know that the lower bounds of confidence interval for variables in the group \{2,5,8\} are zeros, thus, we can conclude the coefficients of the not-true variables are significantly zero. This also demonstrate both BootVS and BLBVS (with different values of $\gamma$) select the variables completely right.
However, both BootVS and BLBVS (even for different $\gamma$)  select the variables completely right.

\subsection{Convergence properties}

In this section, we compare the convergence properties of the two methods.  We first compute the the ground truth of $\xi\{Q_n(P)\}$ based on 1000 realizations of datasets from the underlying distribution $\emph{P}$.
$\xi\{Q_n(P)\}$, stands for  the variance of corresponding variable, and is denoted as  $v_{i,g}$ for the $i$th variable in $g$th group, where $\hat {v}_{i,g}$ is the estimated value and $v_{i,j}$ is the `true' value.
Relative Deviation (RD) of $v_{i,j}$ is used as the comparison criterion: the RD of $v_{i,j}$ from $\hat{v}_{i,j}$ is defined as  $|\hat{v}_{i,g}-v_{i,g}|/v_{i,g}$.  $T_{\mathcal{I}}$ is  the trace of covariance matrix($V_{\mathcal{I}}$) of the grouped variables, $\hat {T}_{\mathcal{I}}$ is the corresponding estimation, and then RD can be written as the following:
\begin{equation}
  RD = \sum\limits_{g \in {\mathcal{I}_g}} {\sum\limits_i {\frac{{\left| {{{\hat v}_{i,g}} - {v_{i,g}}} \right|}}{{{v_{i,g}}}}} }  = \frac{{\left| {{{\hat T}_{\mathcal{I}}} - {T_{\mathcal{I}}}} \right|}}{{{T_{\mathcal{I}}}}}
\end{equation}
where $\hat{T}_{\mathcal{I}} = diag({{\hat V}_{\mathcal{I}}})$;  ${T_{\mathcal{I}}} = diag({V_{\mathcal{I}}})$.

In the process of running the each quality assessment procedure, we record the RD produced after each iteration, as well as the cumulative processing time.  Finally, we can obtain a trajectory of relative deviation versus processing time for each quality assessment procedure.

\begin{figure}[!t]
\begin{center}
\includegraphics[width=0.5\textwidth]{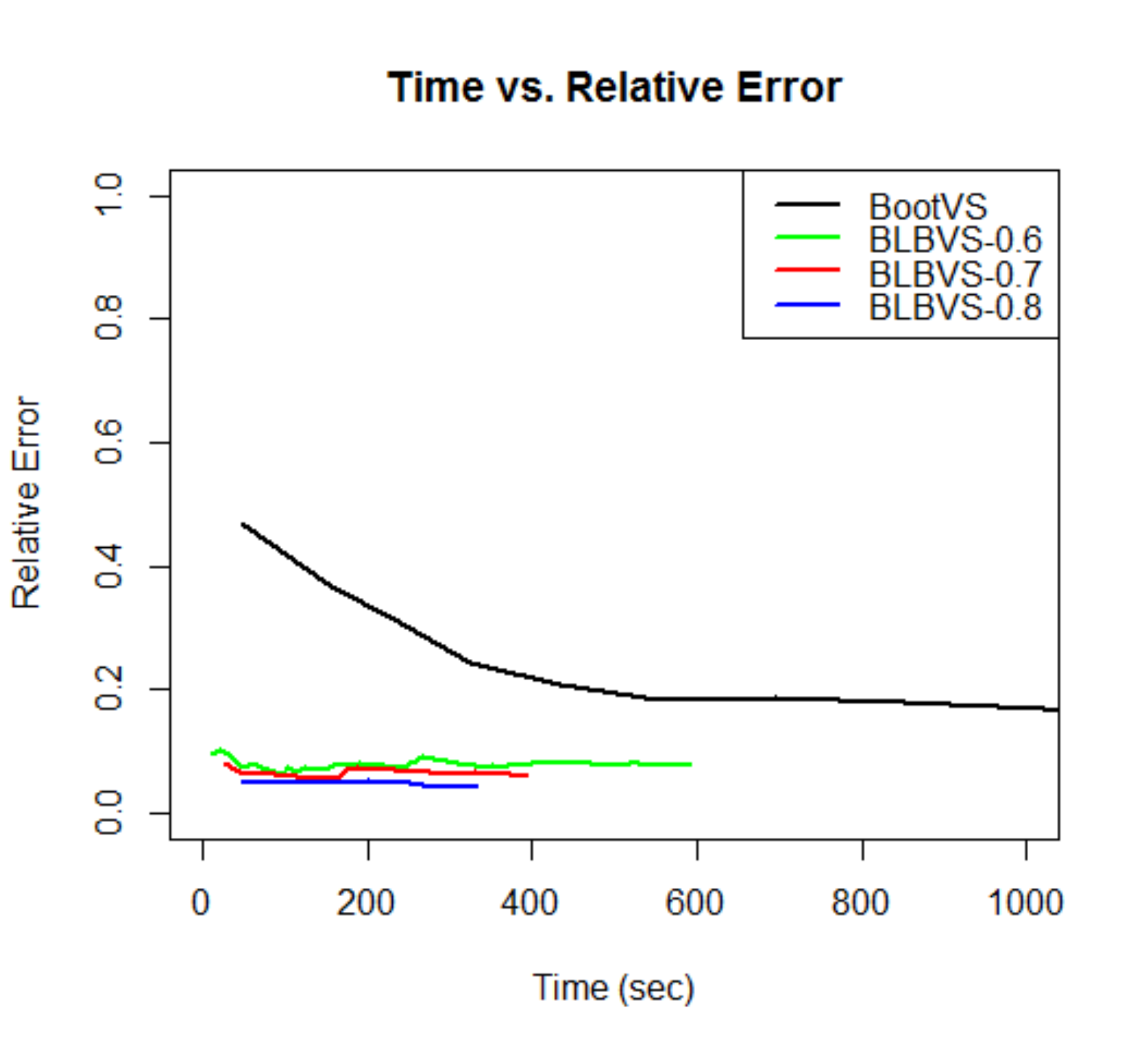}
\end{center}
\caption{RD versus processing time.\ RDs of both methods decrease as processing time increases, while BLBVS decreases more quickly and converge to lower RD than BootVS.  Another good property of BLBVS is that it is robust to various $\gamma$ values.}\label{fig:rele}
\end{figure}

The convergence properties of BootVS and BLBVS (with different $\gamma$) are plotted in Figure \ref{fig:rele}.
We can see that  BLBVS (regardless of different $\gamma$s)  converges to lower RD significantly faster than BootVS does.  For  BLBVS, although the differences are very small for different $\gamma$, we could still recognize that the higher the value of $\gamma$ the lower the RD.

In conclusion, the correctness of selection in both methods are absolutely identical, and the convergence property of BLBVS is significantly better than that of BootVS.  An additional benefit of using BLBVS is that it needs much less time than BootVS to achieve the same accuracy of convergence.

\section{Scalability and computational Analysis}
\label{subsec:scalability}

One noticeable advantage of BLBVS is its characteristic scalable computation  of massive datasets.  Since BLBVS allows parallel and distributed computing, the estimation of different subsets can be calculated by using different individual compute nodes simultaneously, while the traditional method BootVS requires repeated work of using multiple compute nodes for every resample.  BLBVS permits modeling on multiple smaller subsets and resamples in parallel, thus these datasets can be transferred to, stored by, and processed on individual compute node due to their reduced size. In other words, one single compute node can be used for one subset and then intra-node parallelism could be applied across different resamples generated from that subset.   Therefore, comparing with BootVS, BLBVS has better performance on reducing the total computational cost and allowing better application of parallel and distributed computing resources.  In addition, if there is only one compute node available, although it is prohibited to model the observed large dataset by using a single node, it may still be possible to perform variable selection efficiently by processing one subset at a time.

Modern massive datasets often exceed both the processing and storage capabilities of individual processor or compute node, thus the use of parallel and distributed computing architectures is  a popular trend. As a result, the scalability of a quality assessment method closely depends on its ability  to effectively utilizing computing resources. When we use distributed computing to bootstrap resamples directly from a large dataset, it is usually followed by the following process: partitioning data into a cluster of compute nodes, performing estimations across compute nodes simultaneously for each resample, , and computing one resample at a time. This approach, while at least potentially feasible, remains quite problematic. The estimations of each resample require the use of the entire cluster of compute nodes, and bootstrap repeatedly incurs the associated overhead, such as the cost of repeatedly communicating intermediate data among nodes. In such situation, it quickly becomes cost prohibitive to compute many estimates on hundreds of  resamples.

\begin{figure}
\begin{center}
\includegraphics[scale=.3]{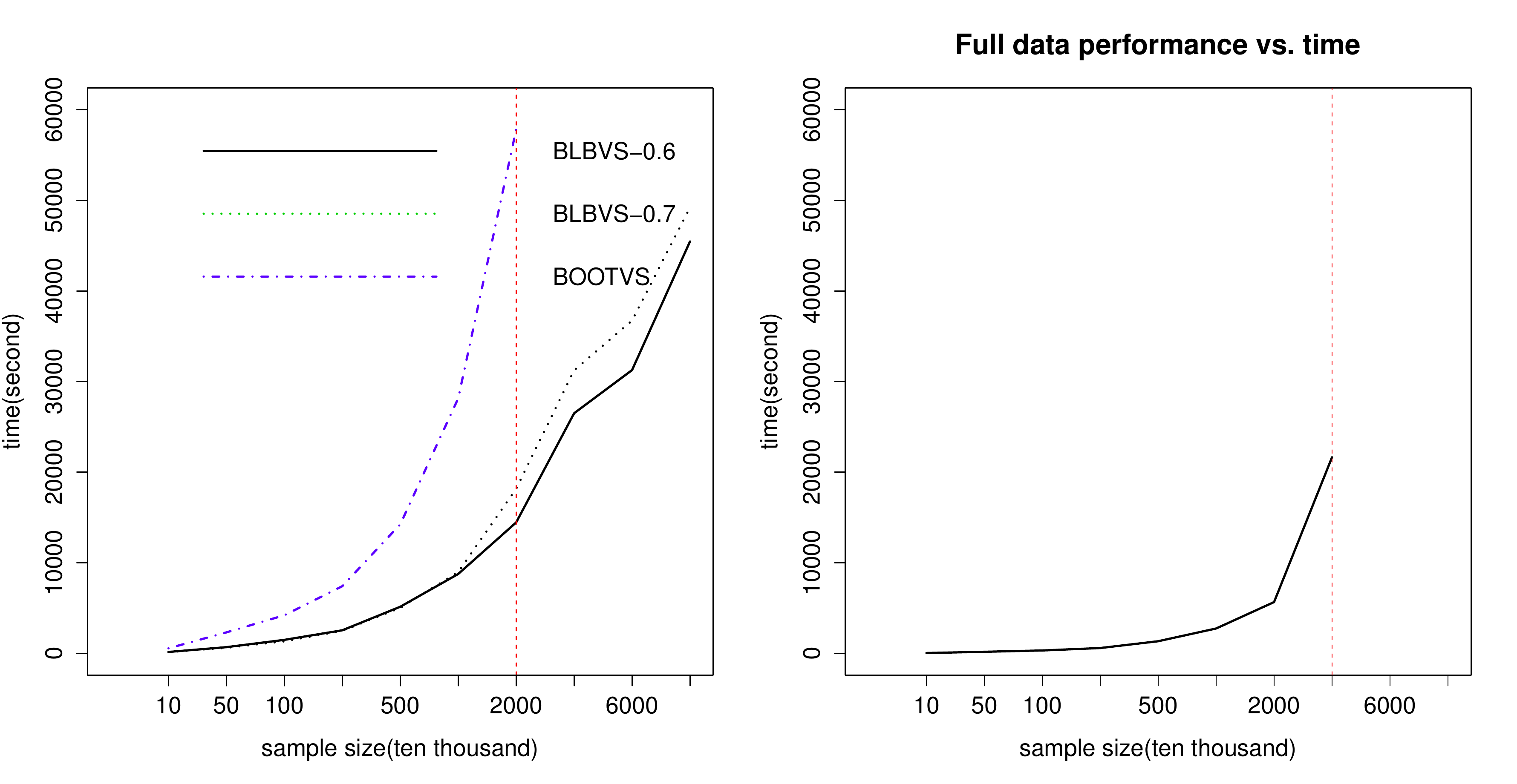}
\end{center}
\caption{Processing time versus size of the original dataset for BLBVS and the BootVS.  The left panel compare the performance of BLBVS and BootVS at the same accuracy.  The right panel shows the processing time of full dataset into the model.}\label{fig:para}
\end{figure}

We now compare the performance of BLBVS and BootVS by simulating a large scale experiment on a distributed computing platform. The model setting of this experiment is specified in section 3. In order to accommodate the large-scale distributed computation, we did some modifications: set $p = 50$  and $n$ in the range of 100,000 to 80,000,000, i.e. the size of the full dataset can be as large as approximately 80 GB. We use a compute machine with a cluster of 4 work nodes, each have 16 GB of memory and 6 CPU (AMD 6344 ) cores; so the total memory of the cluster is 64 GB. The full dataset is partitioned into the four compute nodes. The results of the experiment are shown in  Figure \ref{fig:para}.  The left panel presents the relationship between sample size and the correlated computing time for both methods.  As we can see, when the number of observed dataset reaches 20,000,000, BLBVS uses much less time than that of BootVS to achieve the same accuracy. We also see in the curve of BootVS, the growth rate increases more drastically and the slope gets deeper when the observed dataset size increases.  Meanwhile, BootVS stops working when the size of resample is larger than the memory of the single compute core, as shown in the plot on right, when $n$ reaches 50,000,000, the size of the full dataset is approximate 15GB and larger than the computer memory, and hence BootVS stops working.

\section{Real Data Analysis}

To illustrate the effectiveness of proposed approach on real data, we consider the real dataset of credit card records collected from a commercial bank in Taiwan.  The bank  intended to build  a credit risk prediction model based on basic customer information.  The data size is more than 800 million with about 11 GB of memory, which is overly large to be stored in an ordinary computer memory.  The dataset has 25 categorical predictors (detailed information is specified in Table \ref{table:real data}), and a binary response (0 for without risk and 1 for with risk). For this dataset, we aim to select a set of important predictors which have significant effect on credit risk.
As mentioned in section 2,  owing to the categorical predictors, Group Lasso is used as the penalty for the variable selection method and  the tuning parameter $\lambda$ is chosen by 10-fold cross validation.

As described in Section 4, we use computer servers to solve the problem of computation.
Both BLBVS and  BootVS applied on the real dataset, and we get the same selection results (regardless of different $\gamma$s): $X_6, X_9, X_{14}, X_{15}, X_{18}, X_{19}, X_{23}$.
Looking into these selected important predictors, firstly, it makes sense that a person is more likely to have credit risk if he/she has a record of `forced to stop credit card'. Other key predictors, such as `living area', `housing situation', `education background' and `occupation' illustrate one's wealth and economic status, thus are considered as indirected factors. In addition, `average monthly income for family' and `family economic level', construct borrower financial profile, i.e. the amount of his/her income determines the amount of loan he/she can repay without difficulty.
These selected variables are considered  important factors to asses the borrower's credit in different aspects, and our results can be confirmed from  some existign related studies \protect\cite{liu2004banks,zribi2011factors}.

\begin{table*}[!t]
  \footnotesize
  \caption{Details of the predictors}\label{table:real data}
  \centering
  \begin{tabular}{llcllc}
  \hline
   \textbf{Varibale}&\textbf{Defination}&\textbf{df}&\textbf{Varibale}&\textbf{Defination}&\textbf{df} \\
  \hline
   $X_1$&way of credit card application&7&
   $X_2$&bad record &1\\
   $X_3$&load balance ($\geq$ 8 million yuan)&1&
   $X_4$&refund record&1\\
   $X_5$&refused recored&1&
   $X_6$&force to stop credit card&1\\
   $X_7$&numbers of credit card&3&
   $X_8$&frequency of credit card use&4\\
   $X_9$&living area&3&
   $X_{10}$&urbanization of living area&2\\
   $X_{11}$&gender&1&
   $X_{12}$&age&8\\
   $X_{13}$&marital status&2&
   $X_{14}$&education background&4\\
   $X_{15}$&occupation&20&
   $X_{16}$&average monthly income(individual) &5\\
   $X_{17}$&average monthly expense(individual) &4&
   $X_{18}$&housing situation&5\\
   $X_{19}$&average monthly income(family)&5&
   $X_{20}$&average monthly expense(family)&7\\
   $X_{21}$&religious belief&6&
   $X_{22}$&numbers of living together&7\\
   $X_{23}$&family economic level&4&
   $X_{24}$&blood type&3\\
   $X_{25}$&constellation&11\\
  \hline
  \end{tabular}
\end{table*}

\begin{figure}[!ht]
\setlength{\abovecaptionskip}{-0.5cm}
\begin{center}
\includegraphics[width=0.5\textwidth]{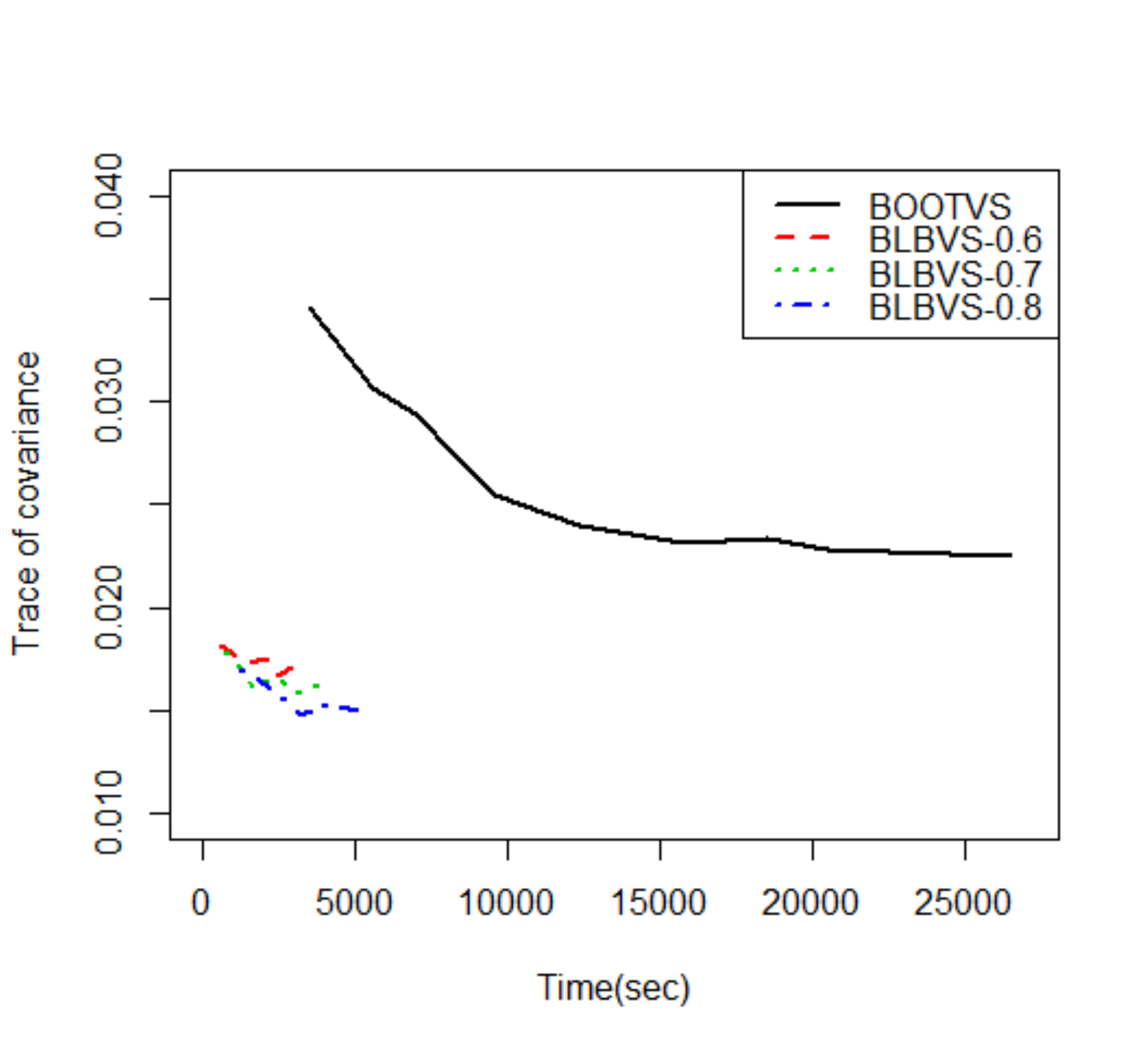}
\end{center}
\caption{Mean standard error versus processing time for the real dataset}\label{fig:realdata}
\end{figure}

Due to the absence of the true underlying distributions of the estimators and the true value of the standard error in a real dataset, it is not possible to evaluate the statistical correctness of any particular estimator quality assessment method objectively.  In stead, we compare the outputs of various methods to address the issue that, without any knowledge of the underlying distribution,  we cannot determine the real covariance of each estimator.
Figure \ref{fig:realdata} shows the convergence properties of BootVS and BLBVS (with different values of $\gamma$) on the real dataset.
As expected, our method (in all values of $\gamma$) not only converges more quickly, it also achieves lower standard error.
And for BLBVS with different $\gamma$, the performances is similar.

\section{Conclusions}
\label{sec:con}

BLBVS is introduced to obtain a computationally efficient means of assessing the quality of estimators and to select significant predictors.
In this article, the method of BLB is referred to separate the original dataset into distinct sub-groups and perform bootstrap resampling at subset level.
 Variable selection is applied at the level of resamples by incorporating the penalty functions, where Lasso is used for GLR and Group Lasso is adopted for GLM.
We have also discussed our method with supportive numerical studies and real data analysis.
Regarding to the numerical studies, the proposed method has excellent performance on both the correctness of selection and convergence. The results of risk forecasting of credit cards shows that it is applicable in the real world data analysis.

One of the remarkable advantages of BLBVS is the better performance on computational profile, and adaptive feature of implementing  modern parallel and distributed computing platforms.
As massive datasets often exceed both the processing and storage capabilities of individual processors or compute nodes, the use of parallel and distributed computing architectures becomes more and more necessary.
In addition, BLBVS possesses generic applicability.
This paper only discusses two kinds of penalties, but based on different situations and model structures, the fundamental form of the model are not fixed thus different penalties can  be chosen. In fact, an excellent aspect of BLBVS is that it can be extended to other variable selection models, such as MCP, Adaptive Lasso, SCAD, etc.

A number of  potential future works remain.
Firstly, to enhance the computational efficiency and automatic nature of BLBVS, a more effective means of adaptively selecting its hyperparameters $\gamma$ is desired.
Secondly, when the dataset is high-dimensional (i.e. tons of predictors), it is necessary to extend BLBVS to the feature screening \protect\cite{xing2001feature}.
Lastly, outliers in variable selection study are commonly encountered \protect\cite{mccann2007robust}, variable selection models that are robust to outliers need further research.

\clearpage
\section*{Acknowledgment}
This study is supported by the Fundamental Research Funds for the Central Universities, and the Research Funds (15XNI011) of Renmin University of China.

\bibliographystyle{plain}
\bibliography{bib-blb}

\end{document}